\documentclass[12pt,a4paper,final]{iopart}

\usepackage{iopams}  
\usepackage{graphicx}
\usepackage[breaklinks=true,colorlinks=true,linkcolor=blue,urlcolor=blue,citecolor=blue]{hyperref}

\begin{document}

\title[Preparing an article for Journal of Statistical Mechanics: Theory and Experiment]{Evolutionary of Online Social Networks Driven by Pareto Wealth Distribution and  Bidirectional Preferential Attachment}

\author{Bin Zhou$^1$, Xiao-Yong Yan$^2$, Xiao-Ke Xu$^3$, Xiao-Ting Xu$^1$, Nianxin Wang$^1$}
\address{$^1$School of Economics and Management, Jiangsu University of Science and Technology, Zhenjiang, 212003, China}
\address{$^2$Systems Science Institute, Beijing Jiaotong University, Beijing 100044, China}
\address{$^3$College of Information and Communication Engineering, Dalian Minzu University, Dalian 116600, China}

\ead{xuxiaoke@foxmail.com}

\begin{abstract}

Understanding of evolutionary mechanism of online social networks is greatly significant for the development of network science. However, present researches on evolutionary mechanism of online social networks are neither deep nor clear enough. In this study, we empirically showed the essential evolution characteristics of Renren online social network. From the perspective of Pareto wealth distribution and bidirectional preferential attachment, the origin of online social network evolution is analyzed and the evolution mechanism of online social networks is explained. Then a novel model is proposed to reproduce the essential evolution characteristics which are consistent with the ones of Renren online social network, and the evolutionary analytical solution to the model is presented. The model can also well predict the ordinary power-law degree distribution. In addition, the universal bowing phenomenon of the degree distribution in many online social networks is explained and predicted by the model. The results suggest that Pareto wealth distribution and bidirectional preferential attachment can play an important role in the evolution process of online social networks and can help us to understand the evolutionary origin of online social networks. The  model has significant implications for dynamic simulation researches of social networks, especially in information diffusion through online communities and infection spreading in real societies.

\end{abstract}


\section{Introduction}

With the rapid development of information technology, online social network platforms have appeared with a novel organizational form that differs from traditional social networks in their ability to bring together users with shared interest or maintained interaction, such as Facebook and Twitter of America, Renren and Tecent QQ of China. Millions of people rely on online social networks to communicate with others where their interactions generate new knowledge~\cite{Faraj_et_al_2011}. Thus the statistics and dynamics of these online social networks are of tremendous importance to researchers who are interested in human behaviors~\cite{Klimek2013Triadic,Ubaldi_et_al_2016}. The systematic research on online social network data has created a new field of network sociology which integrates theories of traditional social networks and modern complex networks, and network science has constituted a fundamental framework for analyzing and modeling complex networks~\cite{Newman_et_al_2011}.

Within the framework of complex networks, studies have concentrated on the structural analysis of online social networks~\cite{Newman_et_al_2016,Lewis_et_al_2008}. Novel network structures of human societies have been revealed. One of the most essential structural characteristic for online social networks is that the degree distribution follows a power-law instead of a normal distribution~\cite{Newman_2003,Barabasi_and_Albert_1999}. Degree distribution is a fundamental quantity measured repeatedly in empirical studies of networks. The degree of an individual is the number of friends that the individual has, and the degree distribution is the fraction for individuals in the network who have exactly the number of friends. In online social networks following a power-law distribution for individual friend count, a few individuals have an extremely high number of relationships to other individuals and most individuals have very few. Among many different theories for generating a power-law distribution, the most prominent explanation is the Barab\'asi and Albert model (BA model) of preferential attachment~\cite{Barabasi_and_Albert_1999}. The BA model is so well established that the preferential attachment is sometimes believed to be the origin of the single power-law distribution.

However, in this paper, our empirical analysis on Renren online social network indicates that the degree distribution is not a single power-law distribution but a two-region power-law distribution in Renren online social network. The similar degree distributions of friend relationship have also been found in other online social networks~\cite{Ugander_et_al_2011}. The community structure and the network connectivity are also important structure characteristics of online social networks, because they play an important role in information diffusion~\cite{Kwon_et_al_2009,Nematzadeh_and_Ferrara_2014} and disease spread~\cite{Salathe_Jones_2010}. The community structure means that the nodes of the network can be easily grouped into sets of nodes such that each set of nodes is densely connected internally. A connected component is a set of individuals among which each pair of individuals are connected by at least one path through the network. The evolutionary characteristics of both the community structure and the network connectivity in Renren online social network are also not consistent with the ones in the simulation network generated by present models~\cite{Klimek2013Triadic,Barabasi_and_Albert_1999, Pacheco_et_al_2006, Toivonen_et_al_2006}. Therefore, the state of art models cannot explain the essential structure characteristics of Renren online social network. A realistic online social network model needs to satisfy the essential structure characteristics of online social networks. Theoretical studies of dynamical processes and collective behavior taking place in online social networks would benefit from the realistic social network models~\cite{lu2011small}. 

How to understand the essential structure characteristics, and whether or not there exist some social mechanisms which result in the essential structure characteristics of online social networks need to be researched deeply. The goal of this paper is to deeply analyze the evolutionary mechanisms of online social networks and propose a model based on the blend mechanisms to reproduce the essential structure characteristics of Renren online social network. We think Pareto wealth distribution and bidirectional preferential attachment can play an important role in the evolution process of online social networks. This paper is structured as follows: firstly, we make the empirical demonstration of Renren online social network and show the essential structure characteristics of Renren online social network. Secondly, we analyze the inherent features of online social networks and propose the social mechanisms to explain the essential structure characteristics. Thirdly, a model is provided based on the blend mechanisms and the evolutionary analytical solution to degree distribution of the model is presented. Fourthly, the simulation is done to reproduce the essential structure characteristics of Renren online social network. Furthermore, the model can also be used to predict both the single power-law degree distribution and the power-law degree distribution with bowing phenomenon in other online social networks, and the mechanism of the bowing phenomenon is uncovered by using the model. Finally, we discuss the significance of the work and conclude with a brief summary of our results.
 
\section{Empirical demonstration and analysis}

The evolution of complex systems in nature and society is from the initial unstable state to the final stable state. The mechanism of driving system evolving plays a dominant role in the evolution process from the initial unstable state to the final stable state. When the system reaches a stable state or dynamic equilibrium, the role of mechanism of driving system evolving will weaken or even disappear. Therefore, analyzing the evolution process from the initial unstable state to the final stable state will help us to explore deeply the mechanism of driving system evolution. The online social network is one of the most complex systems in nature and society. Although there are many online communications that form complex online social networks, detailed topological data is available for only a few, especially for the networks including the evolution process from the initial unstable state to the final stable state. The friend relationship graph of Renren Internet communication represents a well-documented example of the online social network, which is one of the largest online social networks in China. The website of Renren is \url{http://www.renren.com}. The online community is a dynamical evolving one with the new users joining in the community and new connections established between users. Each registered user of Renren has a profile, including his/her list of friends. If we view the users as nodes $V$ and friend relationships as edges $E$, an undirected friendship network $G(V,E)$ can be constructed from Renren. For privacy reasons, the data, logged from 21 November 2005 (the inception day for the Internet community) to 26 February 2006, include only each user's ID and list of friends, and the establishment time for each friend relationship.

\begin{figure}[htbp]
\centering
\includegraphics[width=\linewidth]{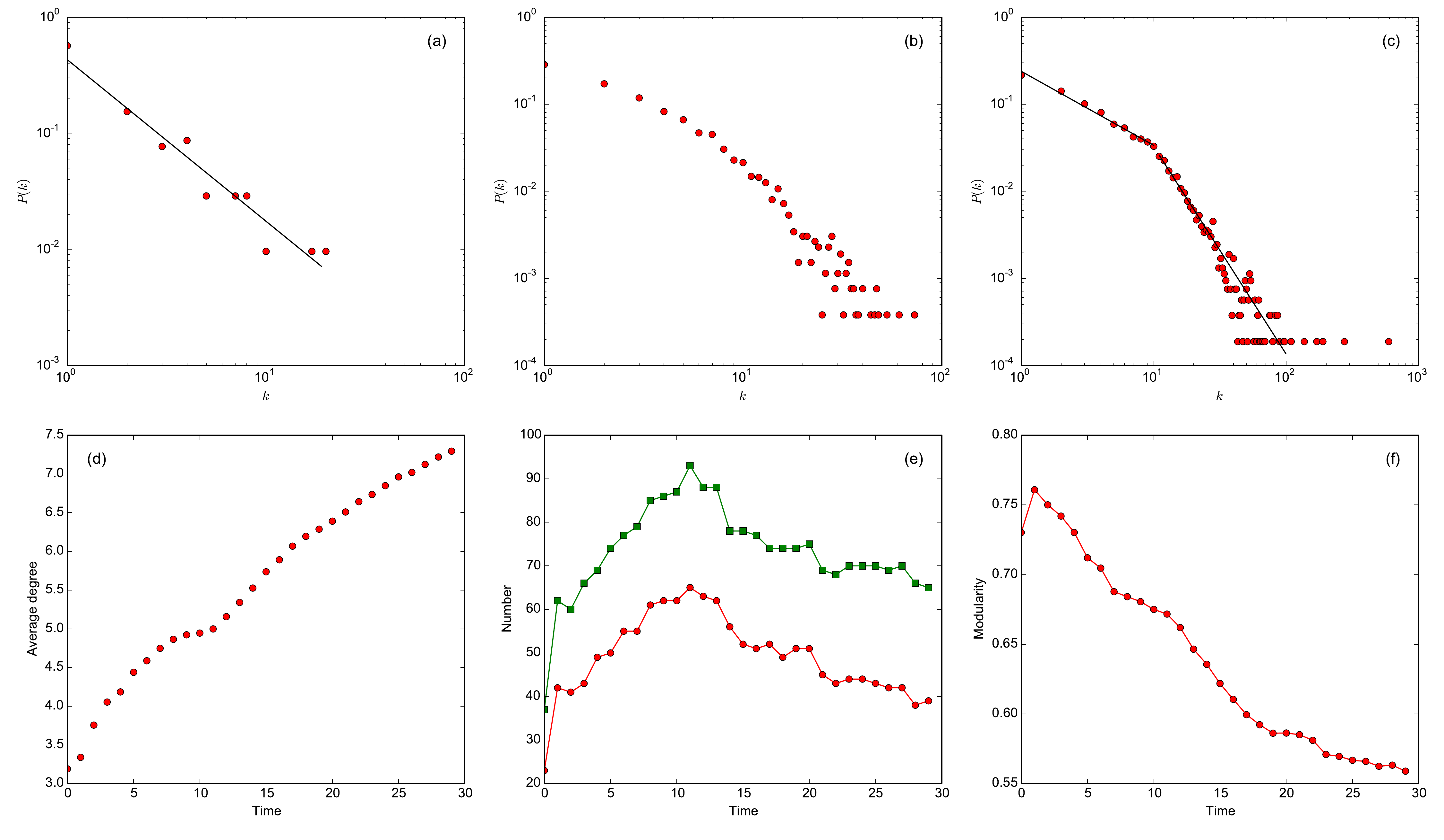}
\caption{The topological evolution of Renren online social network. The evolution time is from 21 November 2005 (the inception day for the Internet community) to 26 February 2006. Figure 1.(a)-Figure 1.(c) are three snapshots of degree distribution chronologically and show the evolution process of degree distribution throughout the initial, middle, and final stages. In Figure 1.(a) and Figure 1.(c), the black solid lines represent the fitting results of empirical data. Figure 1.(d)-Figure 1.(f) contain separately thirty snapshots chronologically extracted from the initial to final stages and show respectively the evolution process of the average degree, community count and connected component count, modularity. In Figure 1.(e), the green square-dash line and the red circle-dash line represent the evolution results of the community count and the connected components count respectively.}
\label{renren}
\end{figure}

Figure \ref{renren}.(a)-Figure \ref{renren}.(c) show chronologically the evolution process of degree distribution in Renren online community in the initial three months. The degree distribution changed from the initial single power-law distribution to the final two-region power-law distribution. The evolution characteristic indicates that some social mechanisms should play a leading role in the evolution process. Figure \ref{renren}.(d) shows the evolution process of the average degree in Renren online social network and the average number of friends for a user increases gradually. The community structure of Renren online social networks is detected using the BGLL method \cite{Blondel_et_al_2008}. Figure \ref{renren}.(e) shows the evolution process of community count and connected component count in Renren online social network. The number of connected components first increases to the peak and then decreases. The evolution tendency of community count almost is consist with the one of the connected component count in the evolution process. We can further quantify this division using the modularity~\cite{Newman_and_Girvan_2004}. The modularity is the fraction of edges within communities minus the expected fraction of edges within communities in a randomized version of the network that preserves the degrees for each individual, but is otherwise random \cite{Molloy_and_Reed_1995}. Figure \ref{renren}.(f) shows that the evolution process of the modularity of Renren online social network. The modularity decreases monotonically in the evolution process. 

\begin{figure}[htbp]
\includegraphics[width=\linewidth]{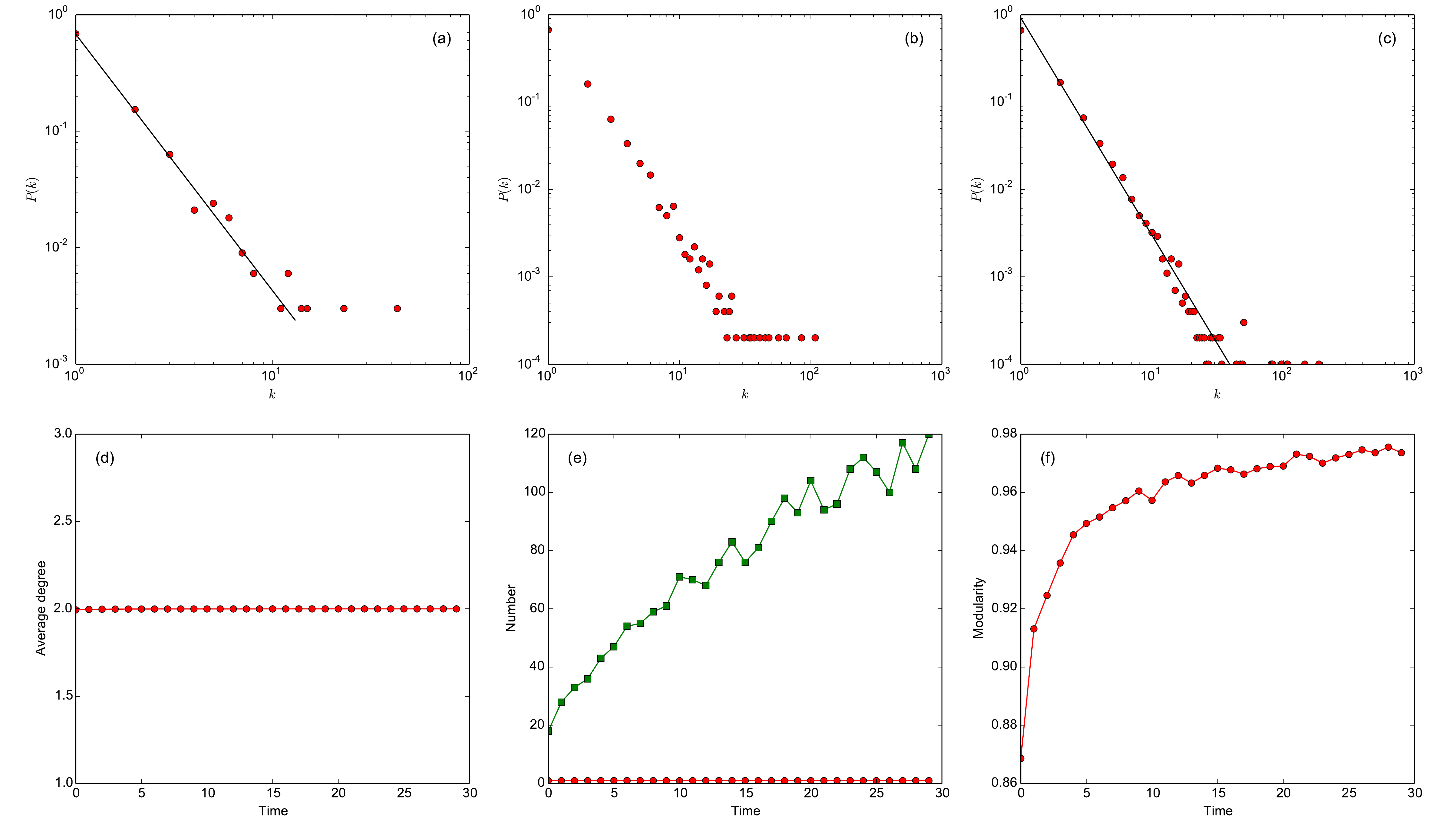}
\caption{The topological evolution of BA model. For BA model, starting with a globally coupled network of three nodes, a new node with one edge is added in every time step and the evolution time is 10000 time steps. Figure 2.(a)-Figure 2.(c) are three snapshots of degree distribution chronologically and show the evolution process of degree distribution throughout the initial, middle, and final stages. In Figure 2.(a) and Figure 2.(c), the black solid lines represent the fitting results of empirical data. Figure 2.(d)-Figure 2.(f)  contain separately thirty snapshots chronologically extracted from the initial to final stages and show respectively the evolution process of the average degree, numbers of community, connected components and modularity. In Figure 2.(e), the green square-dash line and the red circle-dash line represent the evolution results of the community count and the connected components count respectively.}
\label{ba}
\end{figure}

Take the prominent BA model as an example, Figure \ref{ba}.(a)-Figure \ref{ba}.(c) show the evolution process of degree distribution of the simulation network generated by BA model. The degree distribution of BA model always follows the power-law distribution. Due to the rules of BA model~\cite{Barabasi_and_Albert_1999}, the average degree is constant in the evolution process in Figure \ref{ba}.(d). Figure \ref{ba}.(e) shows the evolution process of community count and connected component count in BA model. The number of connected component is one, which indicates that the whole simulation network is connected. The community count increases monotonically in the evolution process. Figure 2.(f) shows that the modularity of the simulation network increases monotonically in its evolution process. By comparing Figure \ref{renren}.(a)-Figure \ref{renren}.(f) with Figure \ref{ba}.(a)-Figure \ref{ba}.(f) respectively, we find that the evolution tendencies of structure characteristics of the simulation network generated by BA model are not consistent with the ones of Renren online social network.

What basic interactions and linking mechanisms result in the essential evolution characteristics of Renren online social network? We will analyze the mechanisms and explain the essential evolution characteristics of Renren online social network. In the evolution process of BA model~\cite{Barabasi_and_Albert_1999}, starting with a small number of nodes $m_0$, at every time step they add a new node with $m$ edges that link the new node to $m$  different nodes already present in the system. To incorporate preferential attachment, they assume the probability that a new node with $m$ edges will be connected to an old node depends on the connectivity of the old node. Aiming at online social networks, we think there are three generic aspects that are not incorporated in BA model.

First, according to the rule of BA model, the relationship between the scale $N$ of simulation network and the evolution time $t$ satisfies the function $N=m_0+mt$. Therefore, the scale of simulation network can grow indefinitely with the time evolution. In contrast, the scale of all online social networks cannot grow indefinitely with the time evolution, because the scale of human in real society is limited. For example, Facebook is built at Harvard University in 2004 and shows the friend relationship between people in the world. At first, the users of Facebook are all the students of Harvard University and the scale of Facebook is only several thousands. With high-speed development of more than ten years, at present, the scale of Facebook has reached to $1.5$ billion, and it has become the largest online social network in the world. However the scale of Facebook is impossible to grow indefinitely, because of the limited number of human being. For the same reason, Renren is one of the largest online social networks in China and also shows the friendship between Chinese people. It is built in 2005 and current scale is about $0.3$ billion. The scale of Renren is also impossible to grow indefinitely. Consequently, a common feature of online social networks is that there is an upper limit on the scale of online social networks and it cannot be more than the potential scale of real society. 

Second, they assume the probability that a new node with edges will be connected to an old node depends on the connectivity of the old node in BA model. Therefore, the process of generating an edge only involve the connectivity information of the old node and not involve other information of the old node in BA model. In online social networks, other information of the users may play an important role in the evolution process of online social networks. The essential evolution characteristics of Renren online social network may be caused by other information of users. The most important one in other information ignored is the social stratification of users in real society. Social stratification is a society's categorization of people into socioeconomic strata, based upon their occupation and income, wealth and social status, or derived power (social and political). As such, stratification is the relative social position of persons within a social group, category, geographic region, or social unit~\cite{Saunders_2006}. An individual class in social stratification has an important influence on the process of establishing his or her social circle. Social stratification of people should play a dominant role in the evolution process of online social networks. Strictly quantitative economic variables are extremely useful for describing social stratification. Wealth variables can vividly illustrate salient variations in the well-being of groups in stratified societies~\cite{Perry-Rivers_2016}. A lot of empirical researches have shown that individual wealth is following the Pareto distribution~\cite{Vermeulen_2017,Levy_and_Solomon_1997,Klass_et_al_2006}. The Pareto distribution, named after the Italian civil engineer, economist, and sociologist Vilfredo Pareto, is a power law probability distribution that is used in description of social, scientific, geophysical, actuarial, and many other types of observable phenomena~\cite{Pareto_1964,Arnold_2015}. An individual wealth class in the Pareto wealth distribution has an important influence on the process of establishing his or her social circle. Therefore, power-law distribution of individual wealth may have an important influence in the evolution process of online social networks. 

Third, we need to analyze the rule of establishing friend relationship between any two users in the evolution process of online social networks. To incorporate preferential attachment in BA model, they assume the probability that a new node with $m$ edges will be connected to an old node depends on the connectivity of the old node. There is a higher probability that the new node will be linked to a node that already has a large number of connections. Therefore, if the new node selected a node by the preferential attachment mechanism, the selected node must be established a connection with the new node in BA model. The preferential attachment mechanism of BA model only involves the demand of the new node and does not involves the demand of the selected node. Consequently, the process of establishing a connection between the new node and the selected node is a unidirectional selection process in BA model. The preferential attachment mechanism of BA model is a unidirectional selection mechanism. However, the unidirectional selection mechanism is unsuitable to explain the process of establishing friend relationship in human social behavior, and many human social behaviors are a universal bidirectional selection process~\cite{Zhou_He_et_al_2014,Zhou_Qin_et_al_2014}. Take Renren as an example, when a user makes a request to another user and wants to establish friendship with another user, the requested user can choose to accept the request or can also choose to reject the request. Only when the requested user accepts the request, the friendship can be established between the two users. Therefore, the process of establishing friend relationship between two users is a bidirectional selection process. Considering the important influence of individual wealth on the evolution process of online social networks, the process of establishing friend relationship between two users should be extremely relevant with the wealth information of the two users. We assume that each user wants to preferentially establish friend relationship with other users which have great wealth, so those users who have more wealth easily establish friend relationship with other users. Compared with the unidirectional referential attachment mechanism of BA model, our bidirectional selection mechanism of establishing friend relationship between two users is a bidirectional preferential attachment mechanism. 

Finally, the mechanisms which include the limited network scale, the power-law distribution of individual wealth and the bidirectional preferential attachment mechanism may be the origin of the essential evolution characteristics of Renren online social network.

\section{Model, analytical solution and simulation}

We build a model based on the blend mechanisms above (see Methods), which can reproduce the observed essential evolution characteristics of Renren online social network. To incorporate the limited scale of simulation social network, we assume there are all $N$ individuals before the individuals begin to establish friend relationship and form a simulation social network, $i,j$ and $l$ are the No. of an individual, and the reasonable ranges of $i,j$ and $l$ are all from $1$ to $N$. The upper limit of the scale of the simulation social network is also $N$. To incorporate the power-law distribution of individual wealth, each individual is assigned a wealth value $\omega$ according to the power-law distribution $p(\omega)=c \times \omega^{-\alpha}$, where $c$ is the normalized constant, and $\omega$ is the an positive integral $\omega \in [a,b]$, $\alpha$ is the power exponent, $a, b$ are both positive integers. To incorporate the bidirectional preferential attachment mechanism, we assume that the probability $\Lambda$ that an individual $i$ will be chosen depends on the wealth $\omega_i$ of the individual, so that $\Lambda(\omega_i)=\omega_i / \sum_{l=1}^{N}\omega_l$. At every time step, two individuals are chosen independently to establish friend relationship and are connected by an edge, and the probability that the two individuals $i$ and $j$ are chosen independently and are connected by an edge depends on the wealth  $\omega_i$ and $\omega_j$  of the two individuals, so that $\Lambda(\omega_i,\omega_j)=\omega_i \times \omega_j / (\sum_{l=1}^{N}\omega_l \times \sum_{l=1}^{N}\omega_l)$. Therefore, there is a higher probability that the friend relationship will be established between two individuals who have great wealth. After $t$ time steps of evolution, the model generates a simulated social network. So far, for the simulation social network for the limited scale, the power-law distribution of individual wealth and the bidirectional preferential attachment mechanism are incorporated in the model.

\begin{figure}[htbp]
\includegraphics[width=\linewidth]{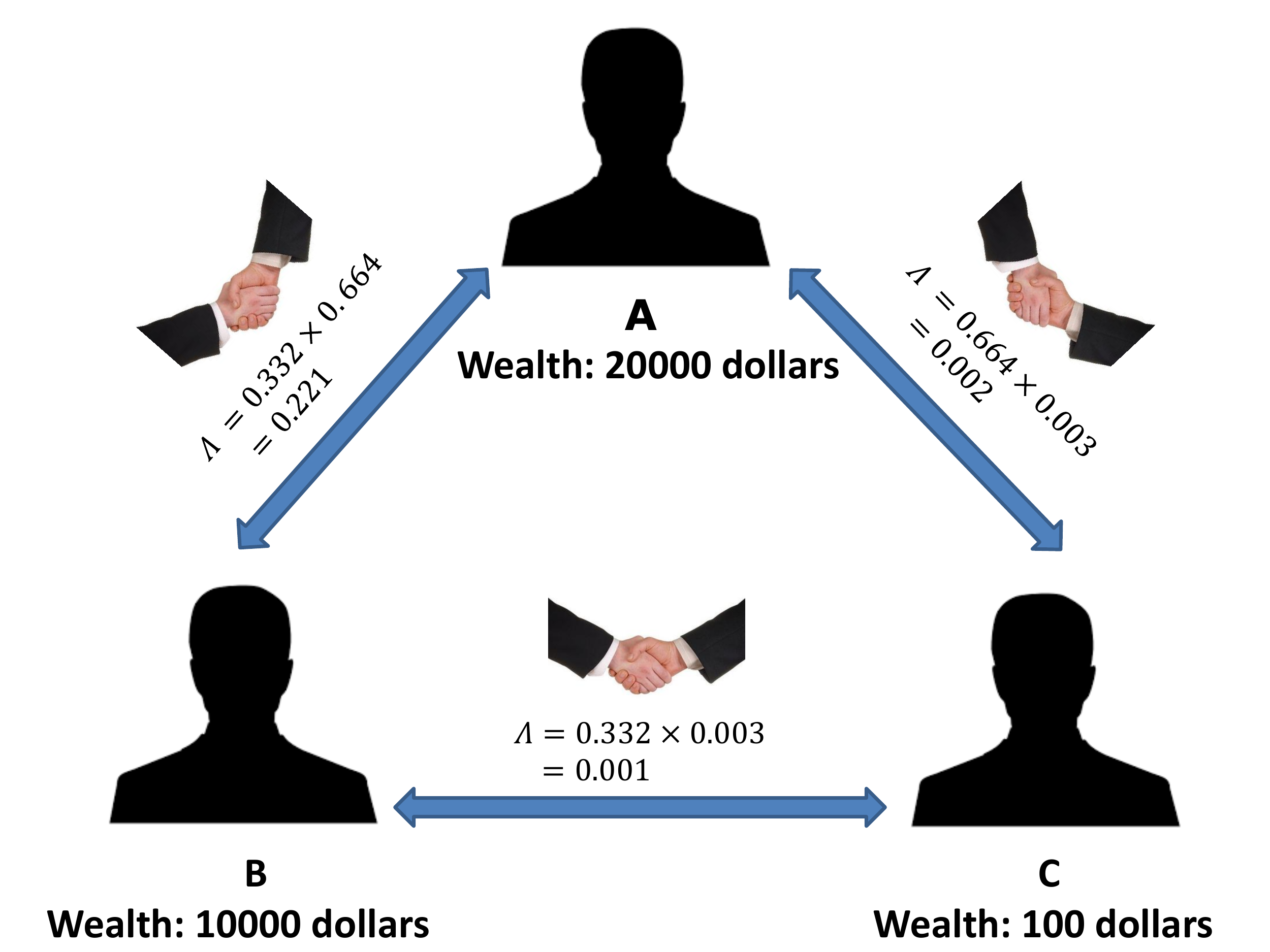}
\caption{The illustration of the proposed model. $\Lambda$ is the probability that the fiend relationship is established between two individuals.}
\label{illustration}
\end{figure}

The illustration Figure \ref{illustration} can help us to well understand the evolution rules of the model. There are three individuals in a social circle, signed as A, B and C. The wealth values of the three individuals is $20000$, $10000$ and $100$ dollars, respectively. The total wealth for them is $30100$ dollars. The wealth proportion of A?OE? B and C in total wealth is $0.664$, and the wealth proportion of B in total wealth is $0.332?OE?0.332 and 0.003 $, respectively. At every time step, two individuals are chosen from the social circle to establish friend relationship. The probability that A and B are chosen to establish friend relationship is $0.221(0.664 \times 0.332)$, and the probability that A and C are chosen to establish friendship is $0.002(0.664 \times 0.003)$, and the probability that B and C are chosen to establish friendship is $0.001(0.332 \times 0.003)$. It only needs less than five time steps $(1/0.221)$ of evolution to make it possible that friend relationship can be established between A and B, but it needs about five hundred or one thousand time steps ($1/0.002$ or $1/0.001)$) of evolution to make it possible that C can establish friend relationship with A or B. Apparently, the probability that the friend relationship is established between A and B is far higher than the one between A and C or the one between B and C.

\begin{figure}[htbp]
\includegraphics[width=\linewidth]{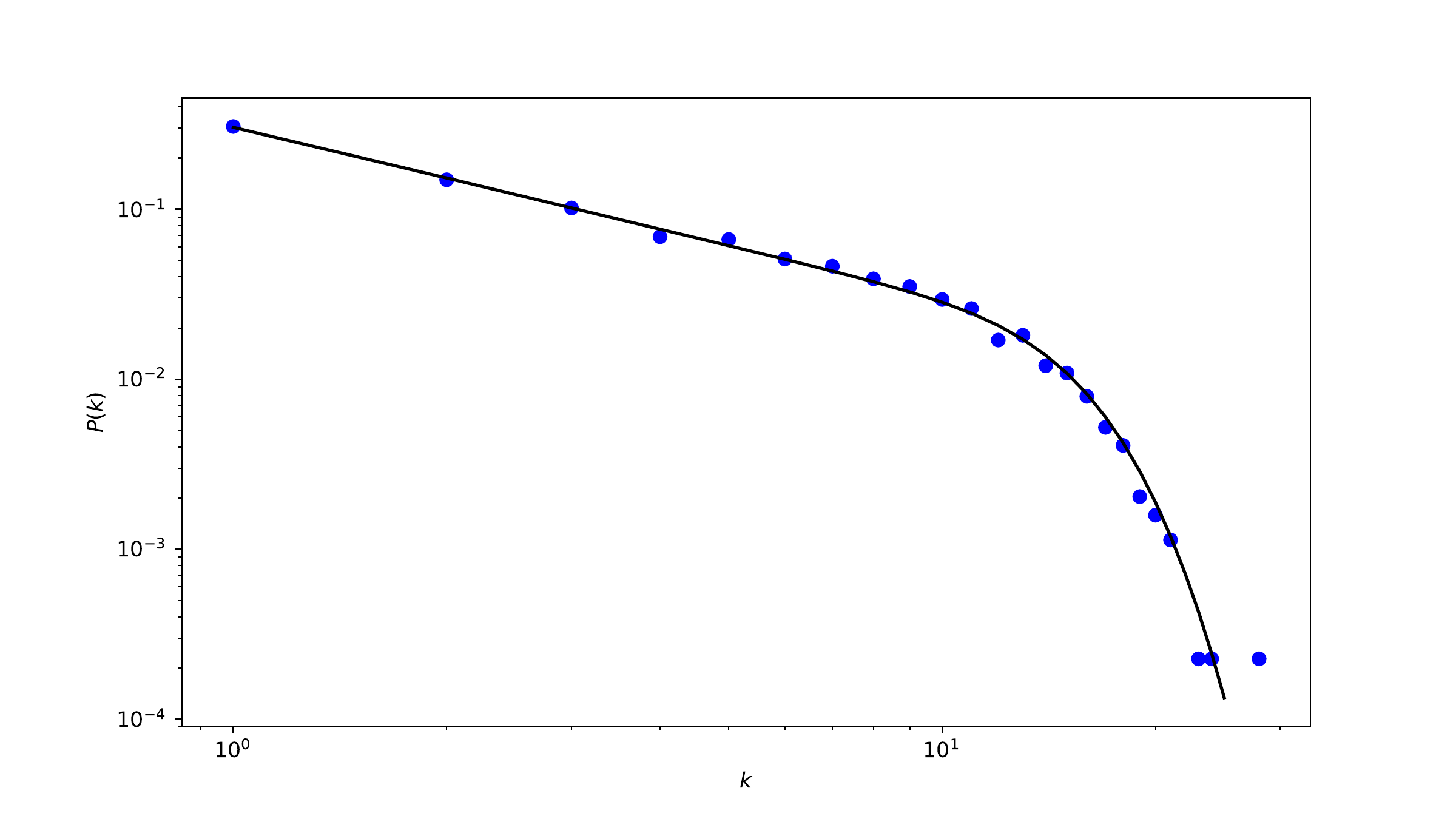}
\caption{Comparison between the simulation results and the analysis results. The parameters are $N=10000, t=10000, a=1, b=1000, \alpha=1$. The blue circle and black solid line represent the simulation results and the analysis results, respectively.}
\label{comparison}
\end{figure}

Then, we will derive the analytical results of the degree distribution evolution in the simulation network generated by the model as follows. In the model, each individual is assigned a wealth value $\omega$ according to the power-law distribution $p(\omega)=c \times \omega^{-\alpha},\omega \in [a,b]$, and the average wealth $\overline{\omega}$ of an individual is $\overline{\omega}=\sum_{\omega=a}^{b}\omega \times c \times \omega^{-\alpha}$. At every time step, two individuals are chosen independently to create one pair of friends and are connected by an edge, and the probability that two individuals $i$ and $j$ are chosen independently and are connected by an edge is $\Lambda(\omega_i,\omega_j)=\omega_i \times \omega_j / (N \times \overline{\omega})^2$. After $t$ time steps of evolution, a simulated social network is generated. In most of real online social networks, their scale is great large and the average degree of an individual is much smaller than the scale. Therefore, most online social networks are spare networks. In the model, when $N$ is great large and $t \ll N^2$, the simulated social network is a spare network. The probability that two different individuals are chosen more than once in $t$ time steps is almost zero and the probability that one individual is chosen twice at one time step is almost zero. Consequently, the duplicate edges and self-connected edges in simulated social network can be ignored. After $t$ time steps of evolution, $t$ edges are established. Once an individual is chosen, the degree of the individual is increased by one. Therefore, the probability $P(k)$ that an individual has $k$ friends after $t$ time steps of evolution is as follows
\begin{equation*}
P(k) = \sum_{\omega=a}^{b} c \times \omega^{-\alpha} \times {t \choose k} \times \left( \frac{2 \times \omega}{N \times \overline{\omega}}\right)^k \times \left(1-\frac{2 \times \omega}{N \times \overline{\omega}}\right)^{t-k},  \qquad  t \ll N^2. 
\end{equation*}
Figure \ref{comparison} shows the comparison of the degree distribution between the analytical solution and the simulation results. The analytical results are in good agreement with the simulation results. That is to say the analytical results are reliable.

\begin{figure}[htbp]
\includegraphics[width=\linewidth]{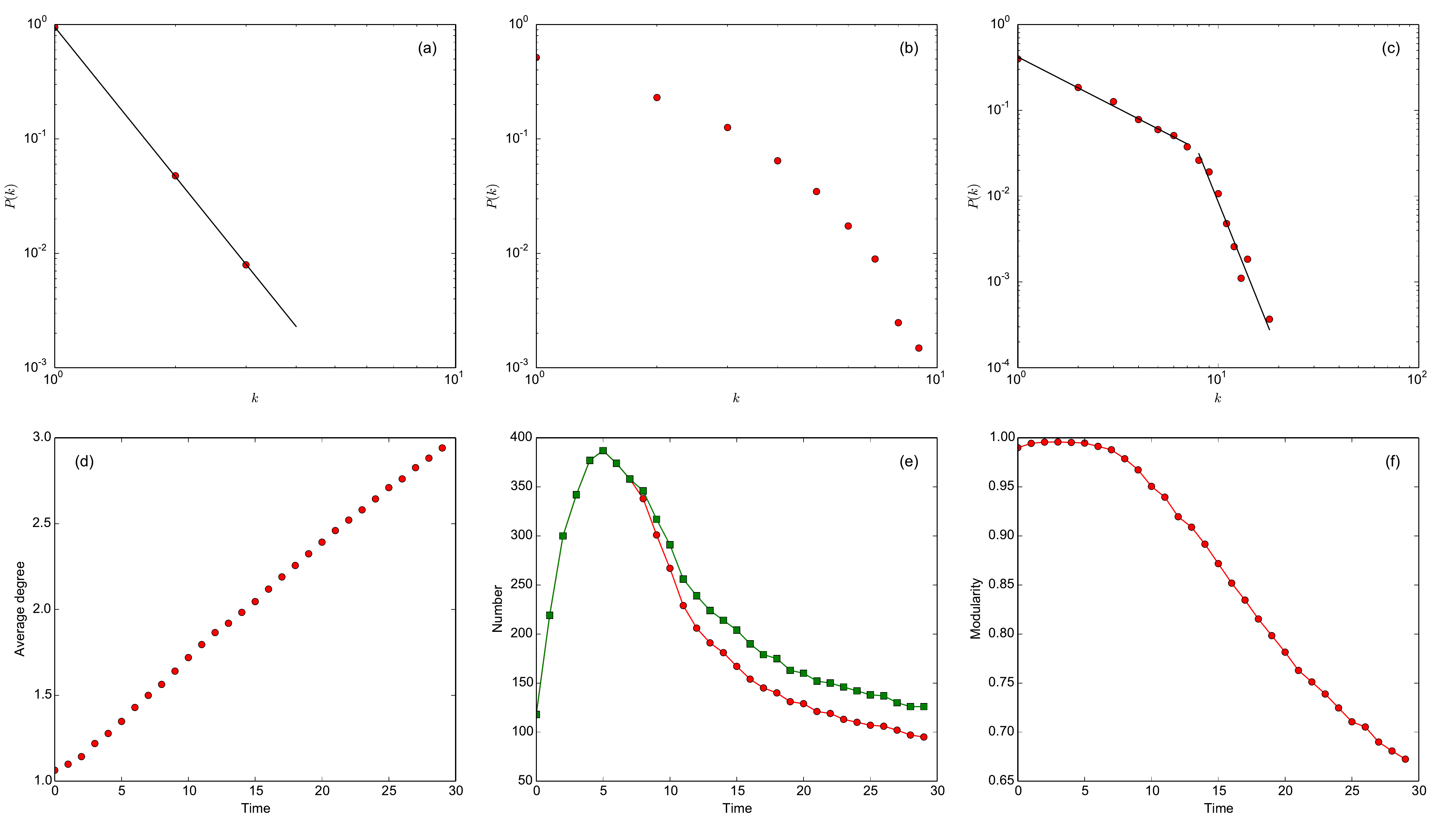}
\caption{The topological evolution of simulation network generated by the model. The parameters of the model are $N=10000, t=4000, a=1, b=10000, \alpha=1$. Figure 5.(a)-Figure 5.(c) are three snapshots of degree distribution chronologically and show the evolution process of degree distribution throughout the initial, middle, and final stages. In Figure 5.(a) and Figure 5.(c), the black solid lines represent the fitting results of simulation data. Figure 5.(d) -Figure 5.(f) contain separately thirty snapshots chronologically extracted from the initial stage to final stage and show respectively the evolution process of the average degree, numbers of community and connected components, modularity. In Figure 5.(e), the green square-dash line and the red circle-dash line represent the evolution results of the community count and the connected components count respectively.}
\label{model_simulation}
\end{figure}

Figure \ref{model_simulation} shows the topological evolution of simulation network generated by the model. Figure \ref{model_simulation}.(a)-Figure \ref{model_simulation}.(c) show chronologically the evolution process of degree distribution. The degree distribution changed from the initial single power-law distribution to the final two-region power-law distribution. Figure \ref{model_simulation}.(d) shows the evolution process of the average degree and the average number of friends for an individual increase gradually. Figure \ref{model_simulation}.(e) shows the evolution process of community count and connected component count. The number of connected components first increases to the peak and then decreases. The evolution tendency of community count almost is consist with the ones of the connected component in the evolution process. Figure \ref{model_simulation}.(f) shows the modularity decreases monotonically in one evolution process. By comparing Figure \ref{renren}.(a)-Figure \ref{renren}.(f) with Figure \ref{model_simulation}.(a)-Figure \ref{model_simulation}.(f) respectively, we find that the evolution tendencies of structure characteristics of the simulation network are consistent with the ones of structure characteristics of Renren online social network. Therefore, a blended mechanism including the limited network scale, the power-law distribution of individual wealth and the bidirectional preferential attachmlent can reproduce the essential evolution characteristics of Renren online social network. The topology structure of the simulation network are similar to the one of Renren online social network. The model can help us to understand the evolutionary origin of online social networks from the perspective of the Pareto wealth distribution and the bidirectional preferential attachment.

\begin{figure}[htbp]
\includegraphics[width=\linewidth]{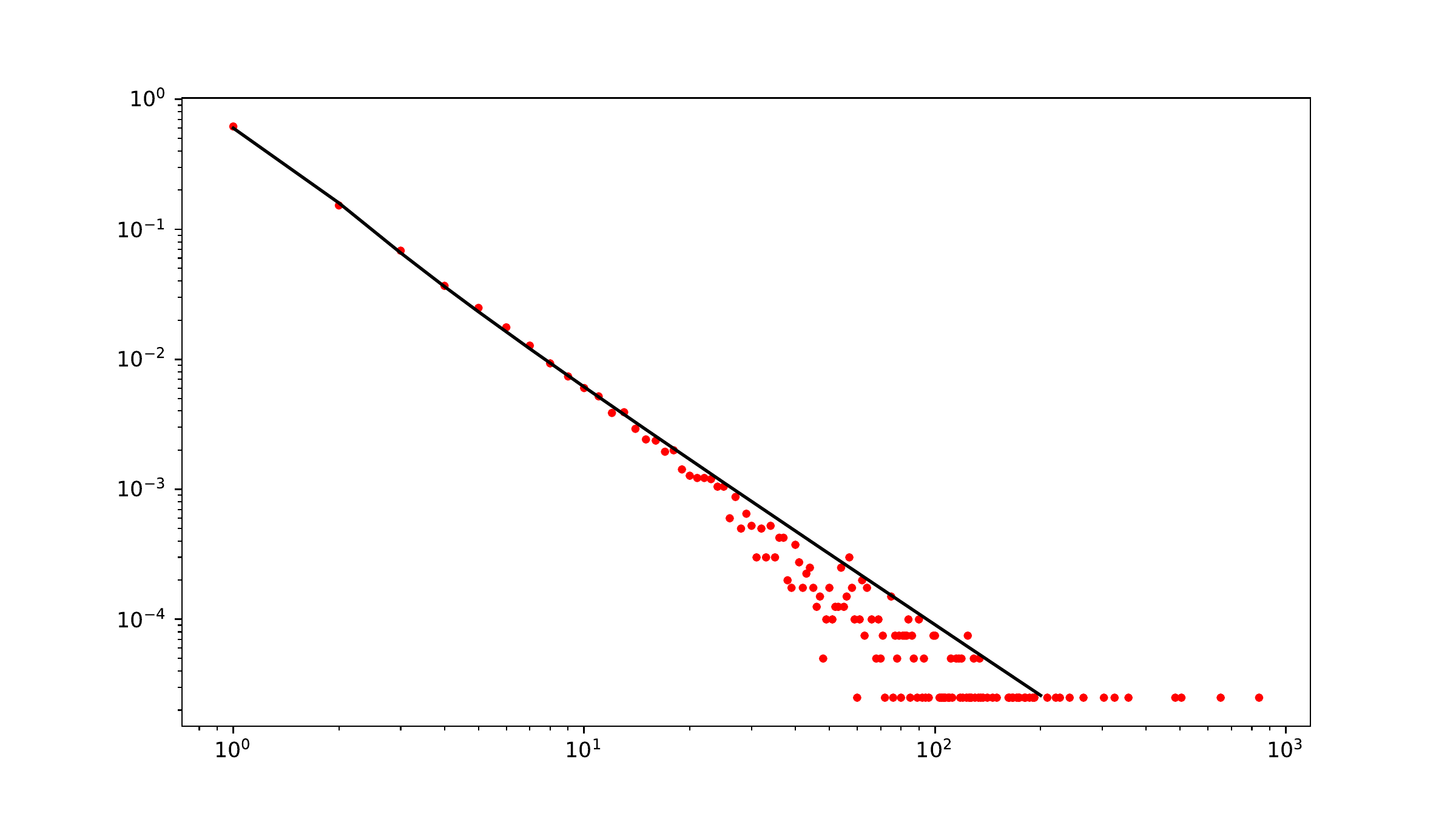}
\caption{Comparison between the predicted results of analytical solution and the experimental results of Flickr. The red circle and black solid line represent the experimental results and the predicted results, respectively. The numbers of nodes and edges of Flickr are $40108$ and $62828$, respectively. The parameters of analytical solution is $N=40108, t=62828, a=1, b=10000, \alpha=1.8$.}
\label{flickr}
\end{figure}

\begin{figure}[htbp]
\includegraphics[width=\linewidth]{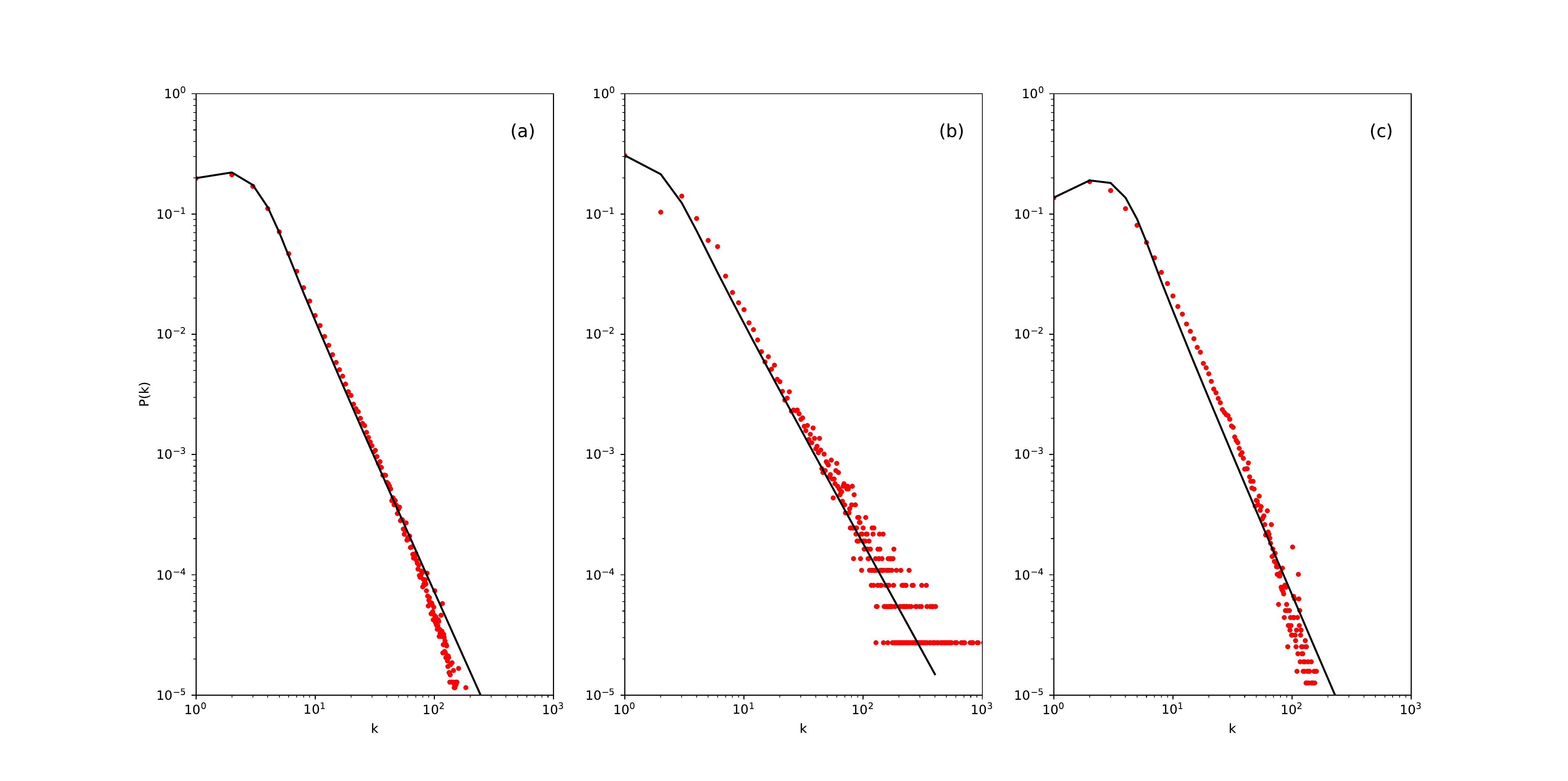}
\caption{The bowing phenomena of three online social networks and the comparison between the predicted results of analytical solution and the experimental results of each online social network. The red circle and black solid line represent the experimental results and the predicted results, respectively. In sub-figure (a), the numbers of nodes and edges in the social network of the academic collaboration relationships are $1560640$ and $4258946$, respectively. The parameters of analytical solution are $N=1560640, t=4258946, a=1, b=200, \alpha=2.2$. In sub-figure (b), the numbers of nodes and edges in the soical network of Enron email communication are $36692$ and $183831$, respectively. The parameters of analytical solution are $N=36692, t=183381, a=1, b=1500, \alpha=1.8$. In sub-figure (c), the numbers of nodes and edges in the co-authorship social network of DBLP computer science are $317080$ and $1049866$, respectively. The parameters of analytical solution are $N=317080, t=1049866, a=1, b=1000, \alpha=2.3$.}
\label{bowing}
\end{figure}

In addtion, the model can well predict the single power-law distribution of online social network. Figure \ref{flickr} shows the comparison of the degree distribution between the predicted results of analytical solution and the experimental results of Flickr. Flickr is an online social network, which is from public data platform \cite{mislove_et_al_2008}, and the website is \url{http://socialnetworks.mpi-sws.org/data-wosn2008.html}. The numbers of nodes and edges in Flickr are substituted into the analytic solution, and the predicted results of the degree distribution are in favorable agree with the empirical results of the degree distribution. In a lot of online social networks~\cite{Garciaet_et_al_2013}, when the degree of a node is very small, the probability of the degree is obviously lower than the one of the power-law distribution, we call these a bowing phenomenon of the degree distribution. Figure \ref{bowing} shows the bowing phenomena of three online social networks and the comparison between the predicted results of analytical solution and the experimental results of each online social network. The three online social networks are about the academic collaboration relationships, Enron email
communication and DBLP computer science co-authorship, which are from public data platform~\cite{Tang_et_al_2008,J2014SNAP}. The website of academic collaboration relationships is \url{https://cn.aminer.org/billboard/aminernetwork}, and the website of both Enron email
communication and DBLP computer science co-authorship is \url{http://snap.stanford.edu/data/}. In three sub-figures, the numbers of nodes and edges in each social network are substituted into the analytical solution, and the predicted results of the analytical solution can also well predict the bowing phenomenon of the degree distribution in each social network.

\begin{figure}[htbp]
\includegraphics[width=\linewidth]{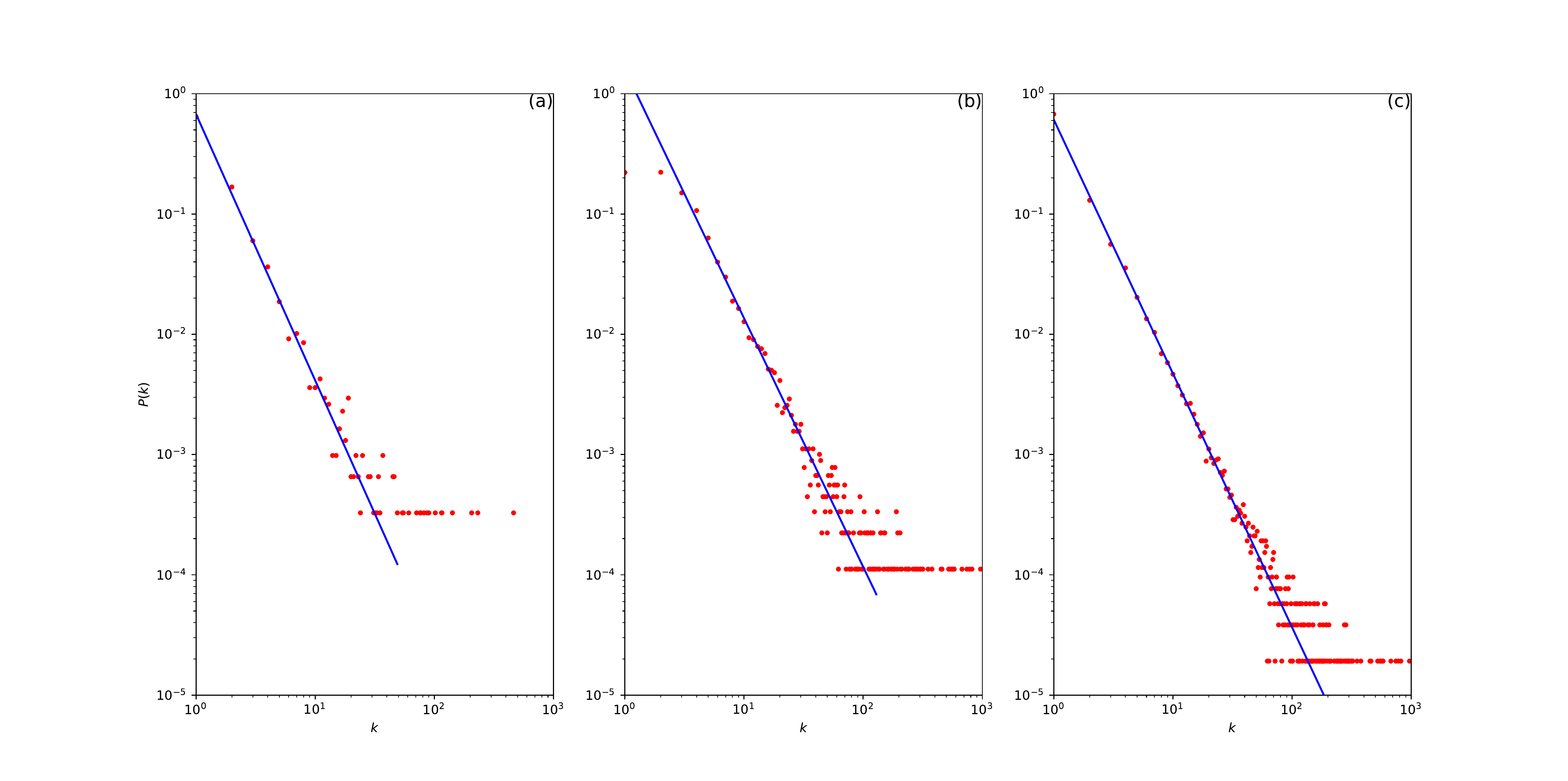}
\caption{\textbf{The simulation process of appearing and disappearing of the bowing phenomenon.} The red circle and blue solid line represent the simulation results and the linear fitting results, respectively. From sub-figure (a) to sub-figure (b), the evolution time $t$ is $5000$ and $50000$ time steps, respectively. Other parameters are constant, $N=10000, a=1, b=10000, \alpha=2$. From sub-figure (b) to sub-figure (c), the network scale  $N$ is $10000$ and $500000$ individuals, respectively. Other parameters are constant, and $t=50000, a=1, b=10000, \alpha=2$.}
\label{bowing_simulation}
\end{figure}

What reason causes the bowing phenomenon of the degree distribution? Our model can account for this. Figure \ref{bowing_simulation} shows the simulation processes of both appearing and disappearing of the bowing phenomenon in the evolutionary simulation network generated by our model. In Figure \ref{bowing_simulation}.(a)-Figure \ref{bowing_simulation}.(b), the evolution time $t$ gradually increased, and other parameters remained unchanged. In Figure \ref{bowing_simulation}.(b)-Figure \ref{bowing_simulation}.(c), the network scale $N$ gradually increased, and other parameters remained unchanged. From Figure \ref{bowing_simulation}.(a) to Figure \ref{bowing_simulation}.(b), it is easy to find that with the time evolution, the degree distribution of simulation network changed from the initial single power-law distribution to a power-law distribution with bowing phenomenon. From Figure \ref{bowing_simulation}.(b) to Figure \ref{bowing_simulation}.(c), it is also easy to find that with the increase of the network scale, the degree distribution of the simulation network changed from the power-law distribution with the bowing phenomenon to the final single power-law distribution. The reason is that in the initial stage of time evolution in Figure \ref{bowing_simulation}.(a)-Figure \ref{bowing_simulation}.(b), there are a lot of individuals with few friends, but with the long time evolution, the individuals with few friends continue to create friend relationships with other individuals so as to have more friends. The number of the individuals with few friends is becoming smaller and smaller, and the bowing phenomenon appears. In Figure \ref{bowing_simulation}.(b)-Figure \ref{bowing_simulation}.(c), compared with the constant parameter $t$, with the increase of the network scale $N$, the time evolution effect is attenuated. Then there are more and more individuals with few friends and the bowing phenomenon disappears. Therefore, according to our model, the interaction of both the evolution time and the network scale leads to the bowing phenomenon of the degree distribution in online social networks. 

\section{Discussion}

Because the topological structure of our social network model is similar to the ones of real-life online social networks, the model of the blended mechanisms has important implications for the dynamic simulation researches of online social networks. The evolutionary characteristics of both community and connected component in the simulation social network are in consistent with the ones in online social networks. These properties play an important role in both information diffusion and infection spreading. The model not only can simulate the two-region power-law degree distribution in Renren online social network but also can well predict the single power-law degree distribution and the power-law degree distribution with bowing phenomenon in other online social networks. Therefore, the blended mechanisms of the model may be a generic property to online social networks and reflect the evolutionary history of online social networks. The Pareto wealth distribution and the bidirectional preferential attachment also may be common mechanisms to others complex networks, such as business networks, biological networks and engineering networks. We expect that the evolutionary mechanism of online social networks can be expanded to other fields of complex networks and help to understand evolutionary origin of other complex systems, with applicability reaching far beyond the quoted examples.

\section{Conclusion}

Firstly, we empirically show the essential evolution characteristics of Renren online social network. From the perspective of the Pareto wealth distribution and the bidirectional preferential attachment, the evolutionary origin of online social networks is analyzed and explained. Then a model is proposed based on the limited network scale, the Pareto wealth distribution and the bidirectional preferential attachment mechanism. Furthermore the evolutionary analytical solution to the model is provided. The simulation results of the model indicate that the model can reproduce the essential evolution characteristics of Renren online social network. Secondly, our model can well predict both the single power-law degree distribution and the power-law degree distribution with bowing phenomenon in other online social networks. We also uncovered the mechanism of bowing phenomenon. Therefore, Pareto wealth distribution and bidirectional preferential attachment can play a dominant role in the evolution process of online social networks, leading from simple individuals to complex online social networks. At last, the model can help us immensely to understand evolutionary origin of online social networks and has great significance in dynamic simulation of online social networks.

\section*{Acknowledgements}

This work is supported by the National Natural Science Foundation of China (Grant Nos. 61503159, 71331003, 71471079) and the Jiangsu philosophy and Social Science Foundation (Grant No. 2016SJB630097 ).

\end{document}